\magnification=\magstep1
\pretolerance 2000
\baselineskip=14pt
\catcode`@=11
\def\vfootnote#1{\insert\footins\bgroup\baselineskip=20pt
  \interlinepenalty\interfootnotelinepenalty
  \splittopskip\ht\strutbox 
  \splitmaxdepth\dp\strutbox \floatingpenalty\@MM
  \leftskip\z@skip \rightskip\z@skip \spaceskip\z@skip \xspaceskip\z@skip
  \textindent{#1}\footstrut\futurelet\next\fo@t}
\skip\footins 20pt plus4pt minus4pt
\def\footstrut{\vbox to2\splittopskip{}}
\catcode`@=12
\def\folio{\ifnum\pageno=0\else\ifnum\pageno<0 \romannumeral-\pageno
\else\number\pageno \fi\fi}
\def\buildurel#1\under#2{\mathrel{\mathop{\kern0pt #2}\limits_{#1}}}
\vglue 24pt plus 12pt minus 12pt
\bigskip
\centerline {\bf  Bounds on the Mobility of Electrons in Weakly Ionized 
Plasmas}
\bigskip
\centerline {A. Rokhlenko, Department of Mathematics}
\centerline {and}
\centerline {Joel L. Lebowitz, Departments of Mathematics and Physics}
\centerline {Rutgers University}
\centerline {New Brunswick, NJ 08903}
\vskip2cm      
\centerline {\bf Abstract}

We obtain exact upper and lower bounds on the steady state drift
velocity, and kinetic energy of electrons, driven by an external field
in a weakly ionized plasma (swarm approximation). The scattering is
assumed to be elastic with simplified velocity dependence of the
collision cross sections.  When the field is large the bounds are
close to each other and to the results obtained from the conventional
approximation of the Boltzmann equation in which one keeps only the
first two terms of a Legendre expansion. The bounds prove rigorously
that it is possible to increase the electron mobility by the addition
of suitably chosen scatterers to the system as predicted by the
Druyvesteyn approximation and found in experiments.
\bigskip
{\bf I. Introduction}
\bigskip
The behavior of the electron mobility in a gas composed of several
species is a subject of continued experimental and theoretical
investigations [1-4]. Of particular interest is the fact that the {\it
addition} of certain types of scatterers, i.e.\ neutral species, to
the gas increases the electron mobility and therefore the electron
current in an applied electric field [3,4].  This effect is
potentially of practical utility and, as was pointed out by Nagpal and
Garscadden [4], can be used to obtain information about scattering
cross sections and level structure of different species.

The fact that the mobility can actually increase with the addition of
scatterers is at first surprising: it is contrary to the well known
Matthiessen rule in metals which states that the total resistivity due
to different types of scatterers is the sum of resistivities due to
each of them [5]. A closer inspection shows that Matthiessen's rule
refers to the linear regime of small electric fields 
while the observations and  analysis in gases [3,4] are in the nonlinear 
high field regime.

This still leaves open the question of the validity of approximations
commonly made in calculating the current of weakly ionized plasmas in
strong fields. We therefore investigate here rigorously the stationary
solutions of the kinetic equation for the electron velocity
distribution function in cases where the electron-neutral (e-n)
collisions are purely elastic and their cross section is modeled by a
simple power dependence on the electron speed. In particular we
establish two-sided bounds for the electron mean energy and drift in
the presence of an external electric field. These bounds show that the
results obtained for the current and energy of the electrons in the
usual approximation, which neglects higher order terms in a Legendre
polynomial expansion and gives the Druyvesteyn-like distribution for
large fields, are qualitatively right and even provide good
quantitative answers. In fact they are sufficiently precise to confirm
an increase in the current for large (but not for small) fields upon
addition of some gases, provided the mass of the added species is
smaller than that of the dominant one, e.g.\ adding Helium to an Xenon
gas, and the different cross sections satisfy certain conditions. We
believe that our analysis can be extended to include more realistic
elastic cross sections and inelastic collisions; these are most
important in practice for enhancement of the electron mobility.
\bigskip
{\bf II. Kinetic Equation}
\bigskip
Our starting point is the commonly used swarm approximation,
applicable to gases with a very small degree of ionization, [6-10]. In
this approximation only e-n collisions are taken into account in the
kinetic equation for the electron distribution function (EDF) $f({\bf
r,v},t)$. The neutrals themselves, which may consist of several
species, are assumed to have a Maxwellian distribution with a
specified common temperature $T_n$. Further simplification is achieved
if the e-n collisions are assumed to be essentially elastic: the
collision integral can then be reduced [1,6] to a differential
operator due to the great difference in the masses of the electrons
and neutrals.  To simplify matters further we consider the case where
the scattering is spherically symmetric.
The stationary kinetic equation for the normalized EDF, in a spatially 
uniform system with constant density $n$ subject to an external electric 
field {\bf $F$}, can then be written in the form [6]$$
-{e\over m}{\bf F}\cdot \nabla_v f=
{1\over v^2}{\partial\over\partial v} \left [\epsilon (v){v^4\over \lambda 
(v)}\left (f_0+{kT_n\over mv}{\partial f_0\over \partial v}
\right ) \right ] +{v\over \lambda (v)}(f_0-f),\eqno(1)$$
$$\lambda (v)=\left [ \sum_{i=1}^SN_i\sigma_i(v)\right ]^{-1},\ \ 
\epsilon (v)=\lambda (v)m\sum_{i=1}^S{N_i\sigma_i (v)\over M_i}.$$
Here $e, m$ are the electron charge and mass, $\sigma_i$ is
the collision cross section with species $i$ whose mass is $M_i$ and number
density is $N_i$, $\lambda$ is the mean free path in the e-n collisions, 
$k$ is Boltzmann's constant,  
$f_0$ is the spherically symmetric part of the distribution function,$$
f_0(v)={1\over 4\pi}\int f({\bf v})d\Omega.$$
We note that $\epsilon$ is a small parameter equal to the ratio
of the electron mass to the mean mass of neutral scatterers, $\epsilon
= m {\bar M^{-1}},$ where ${\bar M^{-1}}=\sum M^{-1}_iN_i\sigma_i/
\sum {N_i\sigma_i}.$
\bigskip
\centerline {\bf A. Velocity independent cross sections}

We shall consider first the case where $\sigma_i(v)$ is independent of $v$
so $\lambda = const$ and $\epsilon =const.$ Taking the electric field parallel 
to the z-axis Eq.(1) can be 
written in the following dimensionless form$$
-E{\partial f\over \partial u_z}=\epsilon{1\over u^2}{\partial\over
\partial u}\left [u^4\left (f_0+{T\over u}{\partial f_0\over 
\partial u}\right )\right ] +u(f_0-f),\eqno(2)$$
where$$
{\bf u}=\gamma {\bf v}, \ \ u=\sqrt {u_x^2+u_y^2+u_z^2},$$ 
$$ \gamma=\sqrt{m\over kT_0 },\ \ 
T ={T_n\over T_0 },\ \ E={e\lambda |{\bf F}|\over kT_0}$$ 
with some fixed $T_0$ specifying the units of the temperature.
We normalize $f$ so that
$$ {1 \over 4\pi} \int f({\bf u}) d^3 u = \int^\infty_0 u^2 f_0 du = 1,
\eqno(3)$$

When $E=0$ the stationary distribution is the Maxwellian with temperature 
$T$, $$
f=f_0=M(u)=\sqrt{2\over \pi T^3} \exp\left ({-u^2\over 2T}\right ),
\eqno(4)$$
$M(u)$ is the unique solution of (2) for $E=0$, $\epsilon \not
=0$. When $E\not =0$ the situation is more complicated. Only for $E$
small compared to $\epsilon$ can we expect the stationary EDF to be
close to $M(u)$.  But in the physically interesting regimes it is
$\epsilon$ which is small compared to $E$. On the other hand if
$\epsilon \simeq 0$ the collisions almost do not change the electron
energy so it is difficult for the electrons to get rid of the energy
they acquire from the field.  The limit $\epsilon \to 0$ is therefore
singular. In particular there is no well defined reference stationary
state for $\epsilon =0$ about which to expand the solution of (2).
\medskip
{\bf B. Legendre Expansion}

The usual method [8] of solving (2) is to expand $f({\bf u})$ in terms of 
the Legendre polynomials $P_l$,$$
f({\bf u})=\sum_{l=0}^{\infty}f_l(u)P_l(cos{\theta}), \quad f_l(u) ={2l+
1\over 4\pi}\int f({\bf u})P_l(\cos\theta)d\Omega_u,\eqno(5)$$
where $\theta$ is the angle between ${\bf u}$ and the field ${\bf
F}$: $cos{\theta}=u_z/u.$ Substituting (5) into (2)  we obtain an infinite 
set of coupled ordinary differential equations for $l\geq 0,\ u\geq 0$. 
These have the form $$
-{E\over 3}\left ({df_1\over du}+{2\over u}f_1\right )=
\epsilon {1\over u^2}{d\over du}\left [u^4\left (f_0+{T \over
u}{df_0\over du}\right )\right ],\ \ l=0,\eqno(6)$$
and $$
E\left [{l\over 2l-1}\left ({df_{l-1}\over du}-{l-1\over u}
f_{l-1}\right )+ {l+1\over 2l+3}\left ({df_{l+1}\over du}+{l+2\over 
u}f_{l+1}\right )\right ]=uf_l,\ \ l=1,2,... .\eqno(7)$$
Eq.(6) can be integrated to give,$$
f_1=-{3\epsilon\over E}u^2\left (f_0+{T\over u}{df_0\over du}
\right ),\eqno(8)$$
where the arbitrary constant of integration was taken to be $0,$ 
using reasonable assumptions on the behavior of $f$ as $u\to 0$ and
$u\to \infty.$

In the conventional [8-10] approximation scheme only two terms of 
expansion (5) are kept. This is equivalent to assuming $f_l(v)\equiv
0$ for $l\geq 2$. One then adds to (8) one more differential 
equation, obtained from (7), for $l=1$ $$
E{df_0\over du}=uf_1.\eqno(9a)$$
Substituting (8) into (9) then yields an equation for $f_0$ $$
\left ( 1+{3\epsilon T\over E^2}u^2\right ){df_0\over du}+
{3\epsilon \over E^2}u^3 f_0=0,\eqno(9b)$$
whose solution is$$
f_0=C \exp\left (-\int_0^u {x^3dx\over T x^2+E^2/3\epsilon }
\right ).\eqno(9c)$$
This $f_0$ becomes the Maxwellian $M(u)$, (4), when $E =0$ and the 
Druyvesteyn [11] distribution $f^D$ when $T = 0$:
$$f_0=f^D = C \exp\left (-{3\epsilon u^4\over 4E^2}\right ),\ \ C=
\sqrt{2}\left ({3\epsilon \over E^2 }\right )^{3/4}{\bigg /}
\Gamma \left ({3\over 4}\right ),\eqno(10a)$$
where $\Gamma$ is the gamma function. Using (9) and (10a) one can 
find $f_1$:$$
f_1=-C{3\epsilon u^2\over E}\exp\left (-{3\epsilon u^4\over 4E^2}
\right ).\eqno(10b)$$
For $T>0$, $f_0$ in (9c) will always have a Maxwellian form for
$u>> (E^2/T\epsilon )^{1/2}.$

The first two harmonics are sufficient to find the mean energy per
particle $W$
and mean speed (drift) $w$ of the electrons which are physically
the most important properties of the stationary state,$$
W={m \over 8\pi} \int v^2 f(v) d^3 {\bf v} = {m\over 2\gamma^2}
\int^\infty _0 u^4f_0du,\ \
w={-1\over 4\pi \gamma^2} \int u_zfd^3u = {-1\over 3\gamma} \int_0 ^\infty 
u^3 f_1 du.\eqno(11)$$
We shall now study the properties of these moments without the
approximations made for explicitly solving Eq.(2). 
\bigskip
{\bf III. Moments of the Distribution Function}
\bigskip
We assume that moments$$ 
{\cal M}_k^{(l)}=\int_0^{\infty}u^kf_l(u)du\eqno(12)$$
exist at least for $0\leq k\leq 9$. Multiplying (7) by a positive 
power $k$ of $u$ and integrating over $u$ yields the equations$$
E\left [-l{l+k-1\over 2l-1}{\cal M}_{k-1}^{(l-1)}+{(l+1)(l+2-k)
\over 2l+3}{\cal M}_{k-1}^{(l+1)}\right ]={\cal M}_{k+1}^{(l)},
\eqno(13)$$
In terms of these moments $w$ and $W$ can be written, using (11) and
(8), as$$
w={\epsilon\over E\gamma}\left [
{\cal M}^{(0)}_5-4T {\cal M}^{(0)}_3\right ],\ \
W={m\over 2 \gamma^2}{\cal M}^{(0)}_4.\eqno
(14)$$
We will now construct estimates of $w$ and $W$ by using 
(8) and (13) to get relations between the ${\cal M}_k^{(0)}$.

\noindent
i) Taking $l=1$ and $k=3$ in (13) and substituting (8) for the
calculation of ${\cal M}_4^{(1)}$ gives$$
{\cal M}_2^{(0)}=1={\epsilon\over E^2}({\cal M}^{(0)}_6-5T {\cal M}^
{(0)}_4).\eqno(15)$$
\noindent
ii) For $l=1,\ k=6,$ Eqs. (13) and (8) yield$$
{\cal M}^{(0)}_5+{1\over 5}{\cal M}^{(2)}_5={\epsilon\over 2E^2}({\cal
M}^{(0)}_9-8T {\cal M}^{(0)}_7).\eqno(16)$$
\noindent
iii) The set $l=2,\ k=4$ allows us to find ${\cal M}^{(2)}_5$:$$
{\cal M}^{(2)}_5=-{10\over 3}E{\cal M}^{(1)}_3=10\epsilon 
({\cal M}^{(0)}_5-4T {\cal M}^{(0)}_3)$$
and eliminate it from (16) to obtain,$$
(1+2\epsilon ){\cal M}^{(0)}_5-8T \epsilon {\cal M}^{(0)}_3=
{\epsilon \over 2E^2}
({\cal M}^{(0)}_9-8T {\cal M}^{(0)}_7).\eqno(17)$$
Further calculation using different $l$ and $k$ will give 
additional equations for the ${\cal M}^{(0)}_j$ which might improve 
the estimates, but we shall use here only (15) and (17).

Exploiting now general bounds on moments of the nonnegative 
density $f_0(u)$ derived in the Appendix we obtain two-sided bounds
for ${\cal M}^{(0)}_3,\ {\cal M}^{(0)}_4,\ {\cal M}^{(0)}_5$, which determine, 
by (14), the electron drift $w$ and mean energy $W$. 
\bigskip
{\bf Inequalities}

The upper bounds on ${\cal M}_j,\ j=3,4,5$, (we have
dropped the superscript zero) can be calculated from (15) using (A5):
$$
{\cal M}_4\leq {\cal M}_6^{1/2} => 1 \geq {
\epsilon \over E^2}\left ({\cal M}_4^2-5T {\cal 
M}_4\right )=> {\cal M}_4^2 
-5T {\cal M}_4-{E^2\over \epsilon }\leq 0.$$
By solving the last inequality one gets$$
{\cal M}_4\leq a,\ \ a={5T\over 2}+\sqrt{{E^2
\over \epsilon}+\left ({5T\over 2}\right )^2}.\eqno(18)$$
The same technique using bounds,$$
{\cal M}_3\leq ({\cal M}_6)^{1/4},\ \ {\cal M}_5\leq
({\cal M}_6)^{3/4} $$
gives$$
{\cal M}_3\leq a^{1/2},\ \ {\cal M}_5 \leq a^{3/2},\ \ {\cal M}_6\leq
a^2,\ \ {{\cal  
M}_6\over {\cal M}_4}\geq a.\eqno(19)$$

The derivation of lower bounds via (15) and (17) is more intricate.
Keeping in mind that $\epsilon$ is small, we use (17) in the
form of an inequality$$
{2E^2\over \epsilon}(1+2\epsilon )>{{\cal M}_9\over {\cal M}_5}-8T
{{\cal M}_7\over {\cal M}_5}\geq 
\sqrt{{\cal M}_9\over {\cal M}_5}
\left (\sqrt{{\cal M}_9\over {\cal M}_5}-8T\right ),$$
where we have used 
${\cal M}_7\leq \sqrt{{\cal M}_5{\cal M}_9}$ in virtue of (A5). 
Using now (A6) with $j=5,\ n=1,\ s=4$ we obtain $$
{{\cal M}_9\over {\cal M}_5}\geq \left ({{\cal M}_6\over {\cal M}_5}
\right )^4$$
and a quadratic inequality for ${\cal M}_6/{\cal M}_5$ whose solution 
is$$
{{\cal M}_6\over {\cal M}_5}\leq b^{1/2},\ \
b=4T+\sqrt{(4T)^2+{2E^2(1+2\epsilon )\over \epsilon}}\eqno(20).$$
We repeat now in (20) the use of (A6) with $i=6,\ k=1,\ s=2$ and 
$i=6,\ k=2,\ s=3/2$ with the results$$
{{\cal M}_6\over {\cal M}_4}\leq b,\ \ \ {{\cal M}_6\over {\cal M}_3}
\leq b^{3/2}.\eqno(21)$$

One can solve (15) for ${\cal M}_6$ in terms of ${\cal
M}_4$ and using (21) obtain the inequality$$
{\cal M}_4={{\cal M}_4\over {\cal M}_6}{\cal
M}_6={{\cal M}_4\over {\cal M}_6}\left (
{E^2\over \epsilon}+5T {\cal M}_4\right )\geq 
b^{-1}\left ({E^2\over \epsilon}+5T {\cal M}_4\right ).$$
Its solution is$$
{\cal M}_4\geq {E^2\over \epsilon (b-5T )}.\eqno(22)$$
Similarly expressing ${\cal M}_5$ and ${\cal M}_3$ through ${\cal M}_5
/{\cal M}_6$ and ${\cal M}_3/{\cal M}_6$ respectively and using
(15), (20)-(22) we find the lower bounds. Together with (19) they 
allow us to write down two-sided bounds
for ${\cal M}_j,\ (j=3,4,5)$ in the form$$
a^{j/2-1}\geq {\cal M}_j\geq b^{j/2-2}{E^2\over 
\epsilon (b-5T )}.\eqno(23)$$ 
These are sufficient, by (14), for the estimation of $w$ and $W$. One 
can write immediately$$ 
{ma\over 2\gamma^2}\geq W\geq {mE^2\over2\gamma^2\epsilon (b-5T)}.
\eqno(24a)$$
Using the definition (14) and the inequality (A5) we obtain $$
{\epsilon \over E\gamma}{\cal M}_5\geq w\geq {\epsilon \over E\gamma}
{\cal M}_5^{1/3}({\cal M}_5^{2/3}-4T),\eqno(24b)$$
which can be combined with (23) for $j=5$ to get explicit bounds on $w$.

The lower bounds in (23) are useless when $E\to 0$ and the solution of (2) 
approaches the Maxwellian. Generally, the inequalities (23) become more 
useful the larger $E$ is.
\bigskip
{\bf IV. Comparison with the Druyvesteyn Approximation}
\bigskip
When the background temperature $T$ is small compared with $E
\epsilon^{-1/2}$ it
can be neglected in (18),(20) and the bounds (24) look simpler:$$
{\epsilon^{1/4}\sqrt{E}\over \gamma}\geq w\geq {\epsilon^{1/4}\sqrt{E}
\over \gamma [2(1+2\epsilon )]^{1/4}},\ \ {mE\over 2\gamma^2
\sqrt{ \epsilon }}\geq W\geq {mE\over 2\gamma^2
\sqrt{2\epsilon (1+2\epsilon )}}.\eqno(25)$$
These bounds specify the electron drift and mean energy
as functions of the electric field and gas parameters within errors of
about $\pm 20\%$ for the mean energy and $\pm 8\%$ for the drift
uniformly in $E$ and $\epsilon$. For comparison $w$ and $W$ 
obtained from the Druyvesteyn distribution [10a] are$$
w\approx 0.897{\epsilon^{1/4}E^{1/2}\over \gamma},\ \ W\approx
0.854{mE\over 2\gamma^2 \sqrt{\epsilon }}\eqno(26)$$
in good agreement with (25) when $\epsilon <<1.$ 

Experimentalists also measure sometimes the transversal $D_t$ and 
longitudinal $D_L$ diffusion constants for the electron swarm. While 
$D_L$ cannot generally be expressed  [2,9] in terms of the velocity
moments,$$
D_t =D={{\bar \lambda } \over 3\gamma}{\cal M}_3$$
is just the isotropic diffusion constant, where ${\bar \lambda}$ is 
the mean free path of electrons (${\bar \lambda}= \lambda$ here). When $T$ 
can be neglected we obtain $$
{\lambda \over 3\gamma}\left [{E^2\over \epsilon}\right ]^{1/4}
\geq D\geq [2(1+2\epsilon )]^{-3/4}{\lambda \over 3\gamma}
\left [{E^2\over \epsilon }\right ]^{1/4}\eqno(27).$$
For comparison $$
D\approx 0.759{\lambda \over 3\gamma}\left ({E^2\over 
\epsilon }\right )^{1/4}$$
in the Druyvesteyn approximation.
\bigskip
{\bf V. Mobility in Binary Mixtures}
\bigskip
The increase of electron mobility $w/F$ in a plasma upon the addition of 
a small amount of a new gas has been observed in [3].   It was calculated 
in [4] within the two-term approximation (8), (9) for
binary mixtures of a heavy noble Ramsauer gas and Helium addition. We
shall show here rigorously that this 
effect exists even with constant collision cross sections.
Using (11) gives$$
w=-{1\over 3\gamma}{\cal M}_3^{(1)}\eqno(28)$$
and for $l=1$ Eq.(13) reads $$
{\cal M}_{k+1}^{(1)}=E\left (-k{\cal M}_{k-1}^{(0)}+2{3-k\over 5}{\cal
M}_{k-1}^{(2)}\right ).\eqno(29)$$
When $E\to 0$ we may neglect the second term in (29) and obtain $$
w\approx {2E\over 3\gamma}{\cal M}_1^{(0)}\approx {4E\over 
3\gamma \sqrt{2\pi T}}={2\over 3}\sqrt{2\over \pi}{eF\lambda \over \sqrt{
mkT_n}},\eqno(30)$$
using (4) and  the initial notation. The resistivity $F/enw$ is here
proportional to $\sum N_i\sigma_i$, which is just Matthiessen's rule. 
 
Let us consider now the case of a strong field, $kT_0 << eF \lambda /
\sqrt{\epsilon},$ for a binary mixture $i=1,2$ and use the two-term ansatz 
(8), (9). We then have the Druyvesteyn distribution (10) with the
moments (26). Using (14) and the notation$$
\alpha ={N_2\over N_1+N_2},\ \ \mu ={M_1\over M_2},\ \ \theta =
{\sigma_2\over \sigma_1}$$
we can write explicit expressions for the drift and mean
electron energy$$
w=0.897\sqrt{eF\over (N_1+N_2)\sigma_1\sqrt{mM_1}}{(1-\alpha +\alpha
\theta \mu)^{1/4}\over (1-\alpha +\alpha \theta)^{3/4}},\eqno(31)$$
$$ W=0.427 {eF\over (N_1+N_2)}\sqrt{M_1\over m}[(1-\alpha +\alpha
\theta)(1-\alpha +\alpha \theta \mu)]^{-1/4}.\eqno(32)$$
Both the current and energy of electrons increase, but the
mobility $w/F$ decreases, as the field $F$ increases.. 

Let us now keep the total gas 
density $N_1+N_2$ constant and vary the relative concentration of 
components by changing $\alpha$. A simple analysis of (31) shows
that $w$ can be non-monotone when both $\theta$ and $\mu$ are larger
than 1. For example, if $\theta =5,\ \mu =20$ then considering $w$ as a
function of $\alpha$, $w = w(\alpha)$, we have 
$${w(\alpha_m)\over w(0)}\approx 1.41,\ \ {w(1)\over w(0)}\approx 0.95$$
Here $w(\alpha_m)$ is the maximum value of $w$ obtained for $\alpha_m 
\approx 0.11$.  The 
drift speed is almost the same in the pure species 1 and 2, but it is
noticeably larger in a mixture. The mean energy of electrons 
changes more: when the lighter component substitutes for the
heavier one it goes down:$$
{W(\alpha_m)\over W(0)}\approx 0.46,\ \ {W(1)\over W(0)}\approx
0.21.$$

There is even a more striking situation, when one just adds the lighter
gas keeping the density $N_1$ of the heavier component constant. In this 
case
$$ w(\delta)\sim {(1+\delta \theta \mu )^{1/4}\over (1+\delta
\theta )^{3/4}},\ \ W(\delta)\sim (1+\delta )^{-1/2}(1+\delta
\theta )^{-1/4}(1+\delta \theta \mu )^{-1/4},\eqno(33)$$
where $\delta =N_2/N_1.$ Increasing $\delta $ we increase the
density of scatterers, but for $\delta = \delta_m =8.5\%$ $$
{w(\delta_m )\over w(0)}\approx 1.4,$$
while the electron energy decreases: $W(\delta_m)\approx 
0.5 W(0).$

We obtained these results approximately - by truncating the series (5).
However comparing (26) with the bounds (24) we see that 
the drift velocity and mean energy for the Druyvesteyn approximation
cannot differ from the exact solution by more than about $+12, -6\%$ and 
$\pm 17\%$ respectively. Hence
the non-monotone dependence of the electron mobility on the
density of the light species holds for the exact solution of the 
kinetic equation (2). When we had $w_{max}\approx 1.40 w(0)$ (within
the approximation) a possible exaggeration of $w_{max}$ by $12\%$ and
underestimation of $w(0)$ at most in $6\%$ could reduce their ratio 
from $1.40$ to $1.16$ but the effect is clearly there without approximations.

The explanation of such unusual behavior of the electron
drift in the nonlinear regime is quite simple. 
When $M_2<M_1$ the addition of species 2 makes the energy transfer from 
the electrons to atoms easier in the elastic collisions. Consequently the 
mean electron energy $W$ will drop
leading to a net increase of the mean free time $\tau (v)\sim \lambda /v.$ 
The competition of $\lambda$ and $v$ is shown by formulas (31)
and (33) where $\alpha , \delta$ represent the concentration of the
lighter species and $\mu$ is proportional to its relative
effectiveness in the energy
transfer.  Adding about $10\%$ of a component with
atoms of mass $m_2\sim 0.05m_1$ the mean
electron energy decreases by about 1/2  implying the increase of $w$ by 
about $40\%$. 

This rise of the electron mobility can be stronger [4] in the case when 
the collision cross section of the main (heavy) component is energy 
dependent and decreases with the electron energy.
\bigskip
\centerline {\bf VI. Simple velocity dependent collision cross sections}

We consider here a one-species plasma with the atoms of mass $M$ and 
generalize the bounds (24) for e-n collision cross section of the form$$
\sigma (v)=\sigma_0\left ({v\over v_0}\right )^p,\eqno(34)$$
where the exponent $p$ can be positive or 
negative in a certain range. Setting$$
v_0^2={eF\over mN\sigma_0},\ \ \  t=\epsilon^{1/2}{kT_n\over mv_0^2},\ \ \
\epsilon ={m\over M},$$
we can rewrite (1) as$$
-\epsilon^{p+2\over 4}{\partial f\over \partial y_z}=\epsilon {1\over y^2}
{d\over dy}\left [y^{p+4}\left ( f_0+{t\over y}{df_0\over dy}\right )
\right ]+y^{p+1}(f_0-f),\eqno(35)$$
where $v=\epsilon^{-1/4}v_0y$ and we have in mind situations with
"strong" electric field $t<<1.$
Using the Legendre series expansion (5) for $f({\bf y})$ we again obtain
the infinite set of coupled equations for harmonics $f_l(y)$ 
$$\epsilon^{-{p+2\over 4}}y^{1+p}f_l={l\over 2l+1}\left ({df_{l-1}\over dy}-
{l-1\over y}f_{l-1}\right )+{l+1\over 2l+3}\left ({df_{l+1}\over dy}+
{l+2\over y}f_{l+1}\right )\eqno(36)$$
for $l=1,2,3,...$ and one more equation$$
f_1=-3\epsilon^{2-p\over 4}y^{p+2}\left (f_0+{t\over y}{df_0\over dy}\right ),
\eqno(37)$$
corresponding to (8). 

Methods similar to those in the Section 2 allow us to derive the 
pair of equations for moments, which generalize (16) and (17):$$
{\cal M}(2p+6)=\epsilon^{p/2}{\cal M}(2),\ \ {\cal M}(3p+9)=
\epsilon^{p/2}c{\cal M}(p+5),\eqno(38)$$
where $$
c={1\over 3}[p+6+4\epsilon (p+3)],\ \ \
{\cal M}(k)=\int_0^{\infty} f_0(y)y^kdy$$
and the background temperature parameter $t$ is neglected for simplicity.
In terms of these moments, which clearly satisfy (A2), we have for the 
electron drift and mean energy$$
w=\epsilon^{1-p\over 4}v_0{\cal M}(p+5),\ \ \ W=\epsilon^{-1/2}{mv_0^2\over
2}{\cal M}(4).\eqno(39)$$

A calculation similar to that described in Section 2 and Appendix 
shows that Eqs(38),(39) yield the following upper (U) and lower (L) bounds 
for for $w$ and $W$:$$ 
w_L\leq w\leq w_U,\ \ W_L\leq W\leq W_U,$$
$$w_L=v_0\left (\epsilon\over c\right )^{p+1\over 2p+4},\ \ 
w_U=v_0\epsilon^{p+1\over 2p+4},\ \ 
W_L={mv_0^2\over 2}\epsilon^{-{1\over p+2}}c^{-{p+1\over p+2}},\ \ 
W_U={mv_0^2\over 2}\epsilon^{-{1\over p+2}},\eqno(40)$$
which give (24) for the velocity independent cross section $p=0$
when $T<<\epsilon^{-1/2}E.$

We can find the approximate solution of (35) $$
f_0^D(y)=C\exp\left [-3\int_0^y{x^{2p+3}dx\over \epsilon^{p/2}+3tx^{2+2p}}
\right ],\eqno(41)$$
using the two-term ansatz which leads to the Druyvesteyn function (9c) for 
$p=0.$ Computing the moments in (39) with the help of (41) yields the 
explicit formulas$$
w_D=\epsilon^{{p+1\over 2p+4}}v_0\left [{2p+4\over 3}\right ]^{p+3\over 2p+4}
\Gamma\left({p+6\over 2p+4}\right ){\bigg /}\Gamma \left({3\over 2p+4}
\right ),$$
$$W_D=\epsilon^{-{1\over p+2}}{mv_0^2\over 2}\left [{2p+4\over 3}\right ]^{1
\over p+2}
\Gamma\left({5\over 2p+4}\right ){\bigg / }\Gamma \left({3\over 2p+4}\right ).
\eqno(42)$$

The bounds in (40) for the drift and energy as functions of the parameter 
$p$ are shown in Fig.1 in the form $w_B/w_D-1,\ W_B/W_D-1$ respectively 
with the Druyvesteyn result (42) for comparison (we use the subscript
"B" for both "L" and "U"). The accuracy of two-term
approximation for our models is quite good.
\bigskip
{\bf Acknowledgments}
\bigskip
This work is supported by the Air Force Office of Scientific Research
Grant No. 95-0159 4-26435.
\bigskip
{\bf Appendix}
\bigskip
The moments ${\cal M}_k$ involved in (15), (17)-(24) are the  integrals 
of the non-negative function $f_0(u)$:$$
f_0(u)={1\over 2}\int_0^{\pi}f({\bf u})sin{\theta}d\theta.$$

We can easily show that $\ln{{\cal M}(k)}$ is a concave function
if one treats $k$ as a continuous variable:$$
{d^2\over dk^2}\ln{\cal M}\geq 0.\eqno(A1)$$
(A1) is equivalent to the inequality$$
{\cal M}{d^2{\cal M}\over dk^2}\geq \left ({d{\cal M}\over dk}\right )^2,
\eqno(A2)$$
which can be written using (12) as$$
\int_0^{\infty}x^k f_0(x)dx\cdot \int_0^{\infty}y^k\ln^2(y) f_0(y)dy-
\left (\int_0^{\infty}x^k\ln{x} f_0(x)dx\right )^2 =$$
$${1\over 2}\int_0^{\infty}\int_0^{\infty}x^k y^k\ln^2\left ({x\over y}
\right )f_0(x) f_0(y) dx dy \geq 0.$$
The concavity implies obviously$$
{\ln{\cal M}_k-\ln{\cal M}_i\over k-i}\leq {\ln{\cal M}_n-\ln{\cal M}_m
\over n-m},\ \ k>i\geq 0,\ n>m\geq i,\ n\geq k.\eqno(A3)$$
Taking $k-i=n-m,\ n-k=j$ we obtain$$
{{\cal M}_k\over {\cal M}_i}\leq {{\cal M}_{k+j}\over {\cal M}_{i+j}},
\ \ k>i,\ j>0.\eqno(A4)$$
For the case $k=m$ (A3) yields inequality$$
({\cal M}_k)^{j-i}\leq ({\cal M}_i)^{j-k}({\cal M}_j)^{k-i},\ \ 0
\leq i<k<j,\eqno(A5)$$
which is equivalent to the following useful set: $$
\left ({{\cal M}_{j+n}\over {\cal M}_j}\right )^s\leq {{\cal M}_{j+sn}
\over {\cal M}_j},\ \ \left ({{\cal M}_i\over {\cal M}_{i-k}}\right
)^s\geq {{\cal M}_i\over {\cal M}_{i-sk}},\eqno(A6)$$
where $i,j,n,k\geq 0,\ s\geq 1$ and $i\geq sk.$
\vfill\eject
\centerline {\bf REFERENCES}
\bigskip
\item {[1]}\ A.von Engel,\ {\it Ionized Gases} (AIP Press, New York, 1993);
pp. 29-30, 243, 292;

C. Brown,\ {\it Basic Data of Plasma Physics} (AIP Press, New York, 1993);

R.Balescu, {\it Transport Processes in Plasmas} (North-Holland,
Amsterdam-Oxford-

New York-Tokyo, 1988).
\bigskip
\item {[2]}\ L.G.H. Huxley and R.W. Crompton, {\it The Diffusion and 
Drift of Electrons in Gases} (Wiley, New York, 1974), Chapter 14.
\bigskip
\item {[3]}\ J.P.England and M.T.Elford, Aust.J.Phys. {\bf 41}, 701 (1988);

M.Kurachi and Y.Nakamura, J.Phys.D: Appl.Phys. {\bf 21}, 602 (1988).
\bigskip
\item {[4]}\ Rajesh Nagpal and Alan Garscadden, Phys.Rev.Lett. {\bf 73},
1598 (1994).
\bigskip 
\item {[5]}\ A.H.Wilson, {\it The Theory of Metals} (Cambridge University
Press, New York, 1953), Chapter 10.
\bigskip
\item {[6]}\ A.V.Rokhlenko and J.L.Lebowitz, Phys.Fluids B {\bf 5}, 1766 
(1993).
\bigskip
\item {[7]}\ N.J.Carron, Phys.Rev. A {\bf 45}, 2499 (1992).
\bigskip
\item {[8]}\ W.P.Allis, Handb.Phys. {\bf 21}, 383 (1956);

Aldo Gilardini, {\it Low Energy Electron Collisions in Gases} (Wiley, New
York, 1972), 

p. 51;

I.P.Shkarofsky, T.N.Johnston, and M.P.Bachynski, {\it The Particle
Kinetics of Plasmas} 

(Addison-Wesley, Reading, MA, 1966).
\bigskip
\item {[9]}\ R.N.Franklin,\ {\it Plasma Phenomena in Gas Discharges}
(Clarendon Press, Oxford, 1976). 
\bigskip
\item {[10]}\ V.E.Golant, A.P.Zhylinsky, and I.E.Sakharov, {\it Fundamentals
of Plasma Physics} (Wiley, New York, 1980).
\bigskip
\item {[11]}\ M.J.Druyvesteyn, Physica {\bf 10}, 61 (1930);

M.J.Druyvesteyn and E.M.Penning, Rev.Mod.Phys. {\bf 12}, 87 (1940).

\vfill\eject
\centerline {\bf Figure caption}
\bigskip
Fig.1. The bounds of the electron drift (Fig.1a) and mean energy (Fig.1b)
as functions of exponent $p$ in (34).

\end